# RECONSTRUCTION OF THE QUASAR BRIGHTNESS PROFILE FROM OBSERVATIONS OF HIGH MAGNIFICATION EVENTS

*Ekaterina Koptelova, Elena Shimanovskaya*

Sternberg Astronomical Institute, Moscow State University

koptelova@xray.sai.msu.ru, eshim@sai.msu.ru

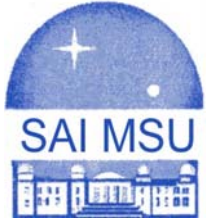
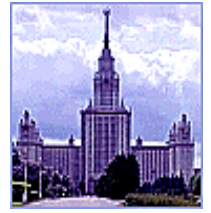

## Abstract

The analysis of the high magnification events in the A and C components of the quadruple gravitational lens QSO2237+0305 observed by OGLE and GLITP collaborations in V band was carried out. The significant light amplifications of the components are interpreted as the effect of microlensing with a fold caustic. For the reconstruction of the one-dimensional source profile the technique based on Tikhonov regularization method was used. The estimates of the effective radius of the quasar's emitting region (the radius within which half of the light is emitted) based on reconstructed profile of the source from microlensing of the A and C components are in the range of 31 and 21 days and correspond to the linear sizes 0.62e+15 cm and 0.42e+15 cm. For the A component the positive crossing of the caustic and for the C component the negative crossing of the caustic was confirmed.

## Introduction

A huge amount of observational data from long-term monitoring programmes of gravitationally lensed quasars now make it possible to study peculiar properties of an individual lens system. Observable brightness variations of gravitationally lensed quasar components reflect both internal quasar processes and microlensing events caused by compact objects located within the lensing galaxy. A microlensing variability of a quasar depends on properties of the source emitting surface and distribution of compact bodies in the lensing galaxy. So, lightcurves of quasar components allow determining the size and the structure of the quasar.

The main difficulty lies in the fact that the magnification distribution due to microlensing is unknown. A magnification map becomes complicated even for two point lenses. In the case of more lenses, magnification distribution can be investigated by means of statistical methods only [1]. When a source crosses a caustic, additional couple of micro images arises (or vanishes) and can be observed as significant magnification of the flux of one of its images up to decimal part of the magnitude. It makes possible to formulate the inverse problem of determination of the size and the structure of the quasar based on observational caustic crossing events [2].

A unique object for the observation of microlensing events is quadruple gravitational lens QSO2237+0305. A time delay between its components is about one day. So, any intrinsic variability of the quasar should manifest in all components almost simultaneously, and uncorrelated brightness variability of the components is due to microlensing. On account of the close galaxy location ($z_l=0.039$), microlensing events (including high magnification events) in this system should occur more often then for the other known gravitational lens systems. By now two such events have been detected – in components A and C of the system – those allow to obtain information about the structure of the quasar using either accretion disk analytical model fitting method [3] or the ill-posed inverse problem regularization method

[4]. Model independent methods of observational data analysis are of special importance, these methods can be used to test adequacy of accretion disk models employed in theoretical studies.

In this work we analyse microlensing high magnification events observed by OGLE and GLITP collaborations and consider the algorithm for reconstruction of the accretion disk brightness profile based on the technique suggested by Grieger et al [5] using Tikhonov regularization approach [6].

## Observations

The gravitational lens QSO2237+0305 has been observed by several collaborations. Since 1997, the long-term monitoring programme of this object has been carrying out by OGLE group. As a result of the observations, two microlensing events with high magnification has been detected in A and C components of the system. The component A rose up to the maximum in the end of 1999, component C reached its peak value in the middle of 1999. At the same time, the flux magnification of the components had been observed by GLITP in R and V bands. The observational data for 1997 – 2000 are published in [7, 8]. Magnitudes of A and C components in V filter, measured by OGLE and GLITP collaborations, have been obtained through Internet. To achieve the denser sampling, OGLE and GLITP data have been combined. The systematic shift of photometric results - 0.065 and 0.13 magnitudes for A and C components respectively – has been taken into account. Then magnitudes have been converted to flux units. Figures 1 and 2 shows fluxes reduced to OGLE photometric system, and error bars.

## Reconstruction of the quasar brightness profile

Let the origin of cartesian coordinates matches the source center, the $x$ axes is perpendicular to a caustic that is considered as a straight line due to small angular size of the source. Let's denote $I(x,y)$ the surface brightness distribution of the source for a distance observer. Crossing the caustic by the source leads to appearance (or vanishing) of an additional couple of microimages. The magnification factor of a point source close to a caustic line along the x axis can be expressed as follows [9]:

$$A(x, x_c) = A_0 + \frac{k}{\sqrt{x_c - x}} H(x_c - x), \qquad (1)$$

where $A_0$ – flux magnification factor associated with microimages created by the lens those don't vanish when caustic crossing (background microimages); $k$ – caustic strength (in units $distance^{1/2}$); $x_c = V_c(t-t_0)$ – the caustic position, $V_c$ – the caustic velocity assumed to be constant, $t_0$ – the point of the caustic crossing by the center of the source ($x=0$); $H(x_c-x)$ – the Heaviside staircase function:

$$H(x) = \begin{cases} 0, & x \le 0 \\ 1, & x > 0 \end{cases}$$

The magnification factor of an extended source with the brightness distribution $I(x,y)$ can be obtained as a sum over infinitesimal sources with their magnification factors:

$$A_{tot}(x_c) = (A_0 + k \iint_{(x<x_c)} \frac{I(x,y)}{\sqrt{x_c - x}} dxdy) / \iint I(x,y) dxdy, \qquad (2)$$

A one-dimensional profile of the source brightness distribution scanned by the caustic line along the $x$ axis is:

$$P(x) = 2 \int_0^{\sqrt{R_s^2 - x^2}} I(x,y) dy, \qquad (3)$$

where $R_s$ – the radius of the source. Caustic crossing is accompanied by drastic magnification of the flux from one of the quasar images. The observable total flux is described by the convolution equation:

$$F_{tot}(x) = \iint_{source} I(x,y) A(x,x_c) dxdy = k \int_x^{x_c} \frac{P(x)}{\sqrt{x_c - x}} dx = A(x,x_c) * P(x), \qquad (4)$$

The flux $F_0$ of other microimages is subtracted from this equation. Let's introduce a regular grid $\{x_i\}$ on the source plane and represent $P(x)$ as a continuous piecewise-linear function:

$$P(x) = P_{j-1} + \frac{P_j - P_{j-1}}{x_j - x_{j-1}}(x - x_{j-1}), \quad x_{j-1} \leq x \leq x_j \tag{5}$$

This allows converting the integral convolution equation into the linear equations set [5]:

$$F_i = \sum_{j=1}^{m} K_{ij} P_j, \tag{6}$$

The matrix $K$ describes the connection between one-dimensional brightness distribution of the source and observations. The direct inversion of this equation is impossible because the matrix $K$ is singular that lead to significant noise amplification. To solve this set of equations, the regularization technique can be applied. To choose the unique solution that makes sense one need to involve additional a priori information such as non-negativity and smoothness of the sought brightness distribution of the source. The main concept of the method is minimization the squared discrepancy between model representation of the microlensing lightcurve and the lightcurve derived from observations with additional penalty term (stabilizer multiplied by regularization parameter α) defining the smoothness of the solution. The smoothing function that needs to be minimized is:

$$M^\alpha[P] = \sum_{i=1}^{m} \frac{1}{\sigma_i^2}(F_i - \sum_{j=1}^{n} K_{ij} P_j)^2 + \alpha \Omega[P], \tag{7}$$

Here the first term represents the squared discrepancy between the model and data, a is the regularization parameter which controls the balance between sought solution and data, W[z] is a stabilizer function through which *a priori* information is introduced into the problem formulation [6]. Generally, a should depend on the input data, the errors, and the method of approximation of the initial problem. One of the way to co-ordinate the regularization parameter with the error of the input information is the discrepancy principle - adoption of a satisfying:

$$\sum_{i=1}^{n}(\sum_{j=1}^{m} K_{ij} P_j^\alpha - F_i)^2 \approx \sigma_{tot}^2, \tag{8}$$

where s$_{tot}$ is the total error of observational data.

Given the source profile is a function that has square integrable second derivatives, one can choose the stabilizer in the following form:

$$\Omega[P] = \sum_{i=2}^{m-1}(P_{i+1} - 2P_i + P_{i-1})^2 \tag{9}$$

The total flux $F_{tot}$ depends on the source properties, the caustic strength $k$ and the contribution of background microimages $F_0$. Values of $k$ and $F_0$ can only be estimated with some uncertainty based on adopted models of mass distribution of the lensing galaxy. The contribution of the background microimages to the total flux can be estimated as an average microimage magnification due to microlensing that is connected with the local dimensionless surface mass density and shear of the lensing galaxy at microimages positions:

$$\langle A \rangle = 1/|(1-\sigma)^2 - \gamma^2| \tag{10}$$

According to the most detailed model of the lensing galaxy 2237+0305 considering its complex structure (buldge and bar), local mass density at A and C positions was estimated as 0.36 and 0.69 respectively, and the shear value was estimated as 0.40 and 0.71. Given these values and the source magnitude in V filter $m_s$~19.11, the background microimages contribution to A and C components was estimated as 0.40 and 0.24 respectively.

Dependency of the caustic strength from the lens parameters has been investigated in [11]. For A and C components the following expressions have been found:

$$\langle k \rangle_A = 0.69 \left( \frac{\langle m \rangle}{\sigma_A} \right)^{1/4}, \qquad \langle k \rangle_C = 0.56 \left( \frac{\langle m \rangle}{\sigma_C} \right)^{1/4}, \tag{11}$$

Given the function of star masses is $\xi(m) \sim m^{-0.3}$ and star masses are in the range $0.1 \leq m \leq 1$ of the Sun mass, then the average star mass in the galaxy is 0.68. So, the values of the caustic strength for two events is 0.81 and 0.56.

## Results

Solution of the inverse problem of the source profile reconstruction with the estimate of the background microimages contribution and the caustic strength has led to a conclusion that the theoretical estimations of the local surface mass density and the shear for the C component is in a good agreement with observational data. The source profile reconstruction for microlensing event in C component was carried out under the assumption about positive (the appearance of a couple of microimages) and negative (vanishing of the couple of microimages) caustic crossing. The discrepancy value in the first case is equal to 85.13, in the second case – 85.04. So, these two possibilities are statistically equivalent. However, the reconstructed profile with the assumptions of the positive caustic crossing contains negative values and significant asymmetry of positive and negative branches of the brightness profile, while the reconstructed profile under the assumption about negative caustic crossing satisfies the flux non-negativity condition.

For the A component, the observed lightcurve is in a good agreement with the model lightcurve, obtained with the assumptions about caustic crossing in the positive direction. The reverse assumption leads to the large discrepancy, reduced chi-square (chi-square divided by the number of free parameters which is equal to 300) is equal to 9.86. Results of the source brightness profile reconstruction for the event in the A component show that theoretical estimates of the background magnification for this components are understated that corresponds to understated value of the surface density or inadequate estimate of shear at the A component position. Best solution was obtained with the background magnification of 0.70, the discrepancy is equal to 103.04. The reconstructed brightness profile is characterized by additional local maxima. Assuming the source brightness distribution is monotonous, additional maxima can be the consequence of unaccounted properties of the magnification distribution in the source plane.

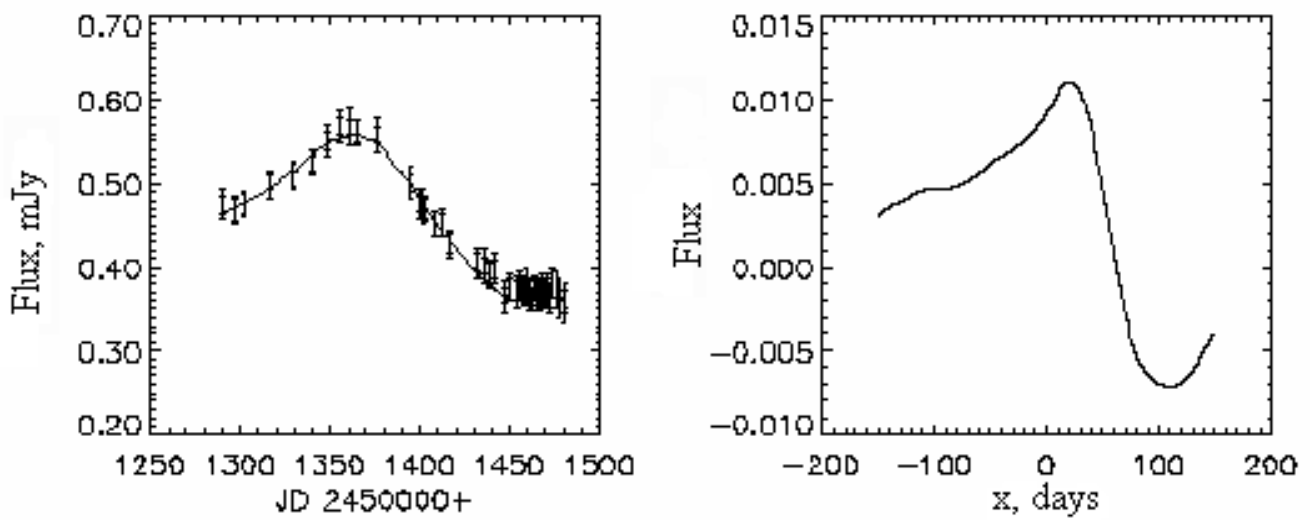

Fig. 1 V band lightcurve of the component C of the gravitational lens QSO2237+0305 from OGLE and GLITP data (left). The solid line represents the lightcurve corresponding to the reconstructed source brightness profile under the assumption of positive caustic crossing (right).

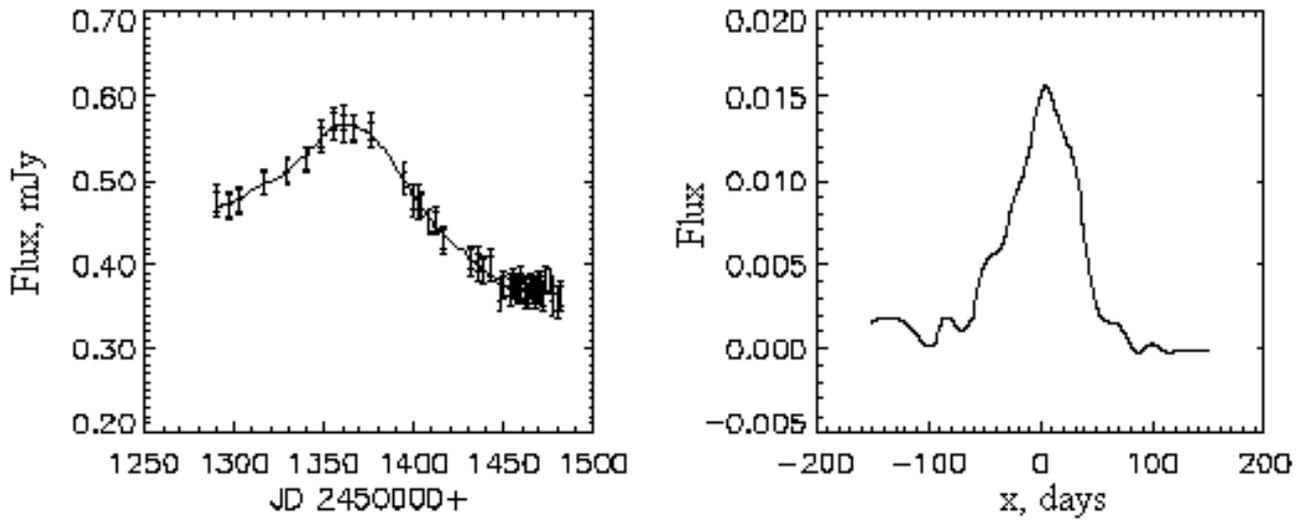

Fig. 2 V band lightcurve of the component C of the gravitational lens QSO2237+0305 from OGLE and GLITP data (left). The solid line represents the lightcurve corresponding to the reconstructed source brightness profile under the assumption of negative caustic crossing (right).

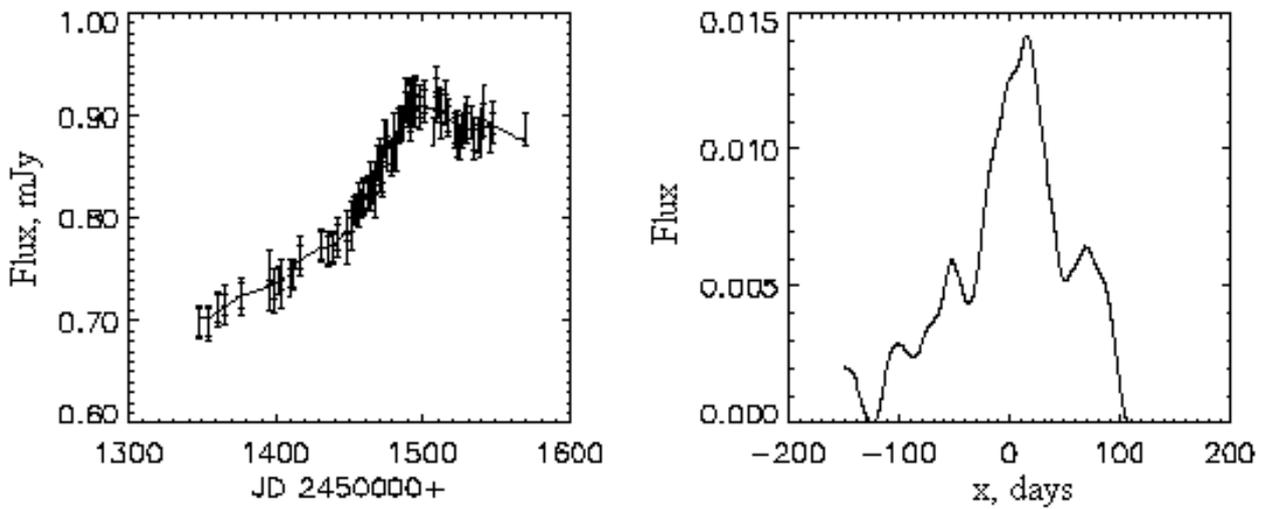

Fig. 3 V band lightcurve of the component A of the gravitational lens QSO2237+0305 from OGLE and GLITP data (left). The solid line represents the lightcurve corresponding to the reconstructed source brightness profile under the assumption of positive caustic crossing (right).

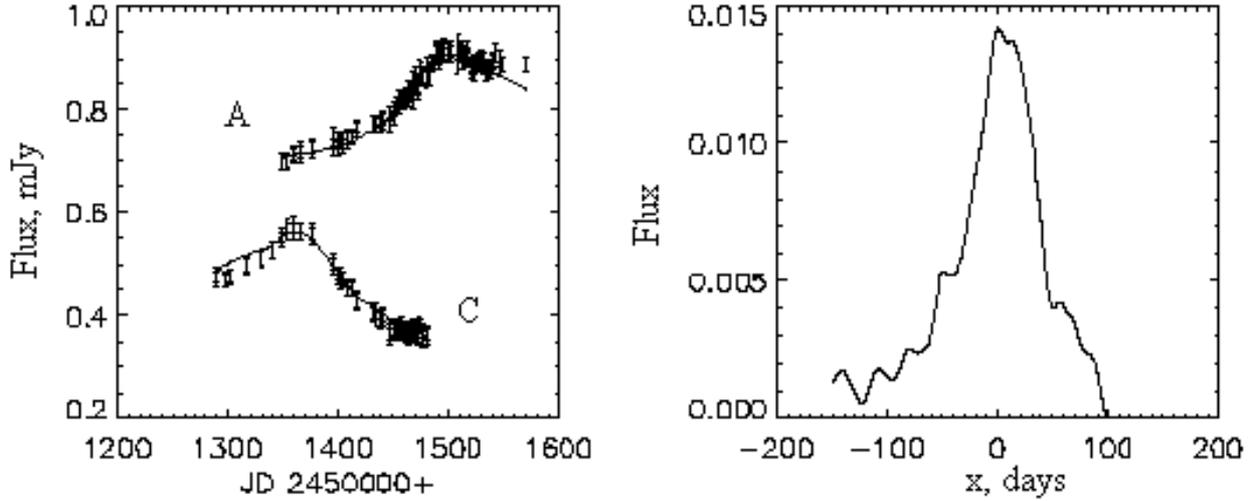

Fig. 4 V band lightcurves of A and C components of the gravitational lens QSO2237+0305 from OGLE and GLITP data (left). The solid line represents the lightcurves corresponding to the reconstructed source brightness profile (right).

**Conclusion**

The resolution of present-day telescopes is insufficient to resolve distant objects such as quasars. However results of theoretical and observational studying of gravitational lenses allows formulating the problem of the quasar brightness profile reconstruction, although under certain assumptions. In this work we used an assumption about crossing the fold by the source. Numerical calculations, and the type of the microlensing lightcurve that allows determining the crossing direction support the assumption. When a macroimage crosses a caustic with rise of additional microimages, a lightcurve has more steep raise with subsequent more gentle slope. A supplementary asymmetry can be concerned with a rotation of an accretion disk and its orientation.

The parameter model of the quasar structure utilizes power dependence of light distribution, dependent on two parameters: source radius and steepness of the brightness profile. Internal radius of the accretion disk can be one more parameter. Taking into account the accretion disk orientation leads to additional parameters, associated with the accretion disk tilt angle to the line of sight and the motion angle to the caustic perpendicular. The accretion disk rotation due to Doppler effect also affects the brightness distribution profile. Because of many factors affecting the source brightness distribution, the problem of the nonparametric quasar brightness profile reconstruction is of particular interest.

Reconstructed brightness profile of the source allows estimating the source size as well. To convert quasar radius estimates from time units to linear units, it is necessary to know projection of the tangential velocity of the caustic to the perpendicular axis. Taking into account angular distances from observer to the lens and to the quasar, the value of the most likely velocity component perpendicular to the caustic line $V_c=300\ km/s$, and the most likely value of the source radius is

$$R_s = V_\tau \Delta t \frac{D_s}{D_l} \approx 2.0 \times 10^{13} \left(\frac{V_c}{300 km/s}\right)\left(\frac{\Delta t}{1 day}\right) cm \qquad (13)$$

The estimates of the effective radius of the quasar emitting region based on the reconstructed brightness profile for high magnification events in the A and C components are 31 and 21 days respectively. It corresponds to linear sizes $0.62 \times 10^{15}$ cm and $0.42 \times 10^{15}$ cm. As can be seen from Figures 1-4, time of the caustic crossing by the source is approximately equal to 100 days, that corresponds to the linear size of the quasar: $2.00 \times 10^{15}$ cm.